\documentclass[lettersize,journal]{IEEEtran}
\usepackage{amsmath,amsfonts,siunitx}
\usepackage{algorithmic}
\usepackage{algorithm}
\usepackage{array}
\usepackage[caption=false,font=normalsize,labelfont=sf,textfont=sf]{subfig}
\usepackage{textcomp}
\usepackage{stfloats}
\usepackage{url}
\usepackage{verbatim}
\usepackage{graphicx}
\usepackage{cite}
\begin{document}

\title{2D Modeling of HTS Coils with $T$-$A$ Formulation: How to Handle Different Coupling Scenarios}

\author{B\'arbara Maria Oliveira Santos, Gabriel Dos Santos, Fr\'ed\'eric Sirois,~\IEEEmembership{Senior Member,~IEEE}, Roberto Brambilla, Rubens de Andrade Junior,~\IEEEmembership{Senior Member,~IEEE}, Felipe Sass, Guilherme Gon{\c c}alves Sotelo,~\IEEEmembership{Senior Member,~IEEE}, Francesco Grilli
         \thanks{This work was financed by the Conselho Nacional de Desenvolvimento Científico e Tecnológico - Brazil (CNPq), and financed in part by the Coordenação de Aperfeiçoamento de Pessoal de Nível Superior - Brazil (CAPES) - Finance code 001.
         B\'arbara Maria Oliveira Santos and  Rubens de Andrade Junior are with Universidade Federal do Rio de Janeiro, Rio de Janeiro, RJ, Brazil.
         Gabriel Dos Santos, Felipe Sass and Guilherme Gon{\c c}alves Sotelo are with Federal Fluminense University, Niter\'oi, RJ, Brazil.
         Fr\'ed\'eric Sirois is with Polytechnique Montr\'eal, Canada.
         Roberto Brambilla was with Ricerca Sistema Energetico, Italy.
         Francesco Grilli is with the Karlsruhe Institute of Technology, Germany.}
\thanks{Corresponding author's e-mail address: francesco.grilli@kit.edu.}
}



\maketitle

\begin{abstract}
Numerical models based on the finite-element method (FEM) are popular tools for investigating the macroscopic electromagnetic behavior of high-temperature superconductor (HTS) applications. This article explains how to use the $T$-$A$ formulation for modeling HTS coils in 2D with different coupling scenarios between the turns. First we consider a racetrack coil  wound from one piece of superconducting tape. Then we consider a coil obtained by winding a cable composed of different HTS tapes.
In the latter case, the tape turns are either electrically connected along their entire length or just at the two ends of the coil: in the model, these two different types of electrical connection are implemented with the help of the electrical circuit module. 
The current density distributions and the AC losses of the coils in the different coupling scenarios are compared and discussed. The limits of applicability of the presented approach are pointed out.
The model is developed for the straight section of racetrack coils, but can be easily adapted to axisymmetric geometries.

 \end{abstract}


\section{Introduction}
\IEEEPARstart{C}{oils} made of HTS coated conductor tapes are being employed in several applications, including magnets and electrical machines. For the design of these applications, the estimation of the electromagnetic behavior of the HTS coils (for example, their power dissipation when they carry time-varying currents) is very important and usually requires dedicated numerical models. Those models are often based on  finite-element analysis.

The $T$-$A$ formulation, originally proposed in~\cite{Zhang:SST17}, has emerged as a popular approach to efficiently and accurately simulate the electromagnetic behavior of devices made of HTS coated conductors~\cite{Huber:SST22}.
The main appeal of this formulation is that, by assuming that the superconducting layer can be considered as an infinitely thin sheet~\cite{Brandt:PRB93,Zeldov:PRB94}, it leads to short computation times, especially for 2D problems where only the transverse cross section of superconductors is encountered.

\begin{figure}[h!]
     \centering
    \includegraphics[width=8 cm]{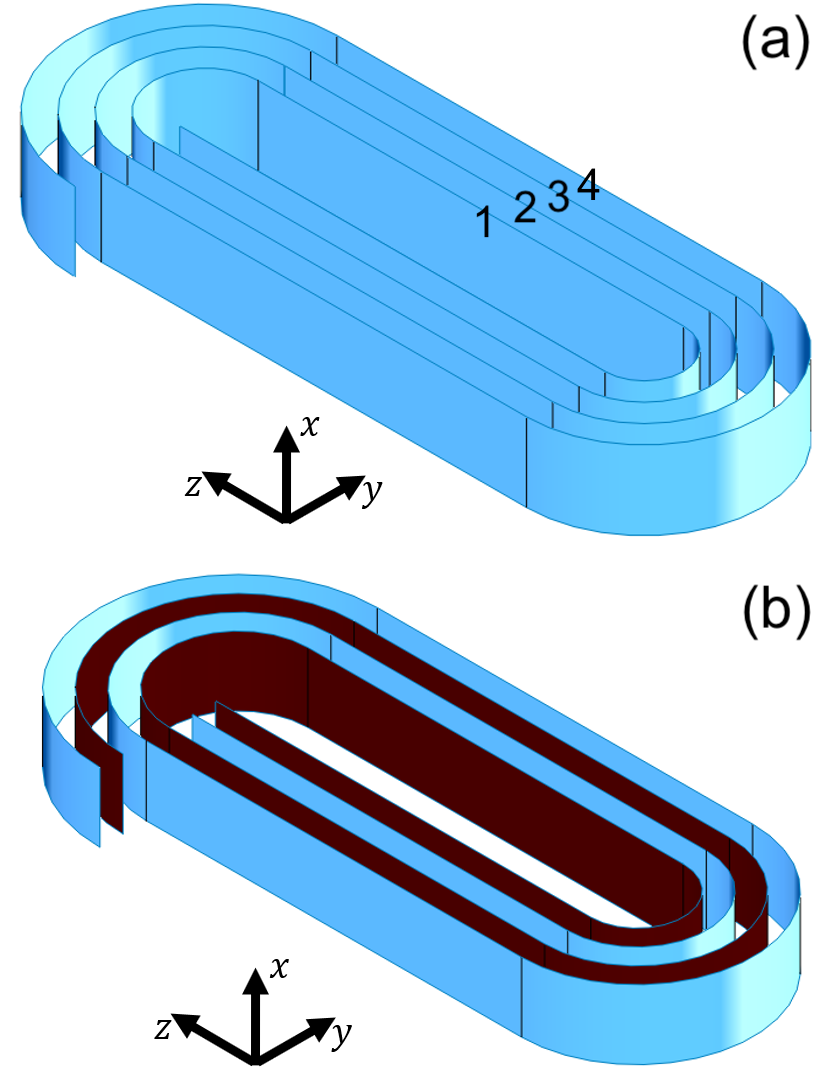}
     \caption{\label{fig:Gabriel}(a) Racetrack coil wound from one piece of tape. (b) Racetrack coil wound from a cable made of two stacked tapes. In the situation represented in (b), the stacked tapes can be electrically connected along their entire length (coupled case) or only at the coil's ends (coupled-at-ends case).}
\end{figure}

When simulating coils made of HTS coated conductors, different scenarios for the electrical connection between the turns can be considered:
\begin{enumerate}
\item The coil is wound from one piece of tape. We refer to this situation as {\it uncoupled} tape turns, because when one considers a 2D cross section of the coil, the tape turns behave as independent conductors positioned side by side, each carrying the same current -- Fig.~\ref{fig:Gabriel}(a).
\item The coil is wound from a `cable' made by stacking several HTS tapes on top of each other --  Fig.~\ref{fig:Gabriel}(b). Within each cable turn, the tapes can be electrically connected along the entire length of the coil or just at its ends. We refer to these cases as {\it coupled} and {\it coupled-at-ends} tape turns, respectively.
\end{enumerate}

From the point of view of modeling, case 1) can be simulated with the $T$-$A$ formulation as in its original implementation~\cite{Zhang:SST17,Liang:JAP17}, because the current flowing in each simulated tape is known (in particular, it is the same).
On the other hand, in case 2) the current flowing in each cable turn is known, but the current flowing in each tape of the cable is not. In this contribution, we couple the $T$-$A$ formulation to an electric circuit model in order to determine the current flowing in each tape of the cable for case 2), both in the coupled and coupled-at-ends situations.

All these cases have been analyzed in~\cite{Pardo:TAS19} with the minimum electro-magnetic entropy production (MEMEP) method. Here, we describe how this can be done with the $T$-$A$ formulation of Maxwell's equations, implemented in the commercial FEM software package COMSOL Multiphysics. 
One of the advantages of the $T$-$A$ formulation is that it provides a practical way of modeling superconducting electrical machines in 2D: the electromagnetic modeling of the whole machine and the calculation of the current and field distributions inside superconducting tapes can be done simultaneously, within the same simulation environment~\cite{Benkel:TAS20}.

\section{Current constraints for Different Coupling Scenarios}
This section presents the settings of the current constraints for 2D problems in the various coupling scenarios. In all cases, the tape turns are modeled as infinitely thin shells (i.e. lines in the 2D problems considered here).

In general, the current flowing in a given conductor $i$ can be imposed by constraining the current vector potential $\mathbf T$ as follows:
\begin{equation}\label{eq:circulation}
I_i=\iint_\Omega {\mathbf J}\,{\rm d}\Omega=\iint_\Omega \nabla \times {\mathbf T}\,{\rm d}\Omega=\oint_{\partial\Omega}T_t \,{\rm d}\ell,
\end{equation}
where $\Omega$ is the 2D cross section of the tape, $\partial\Omega$ is its contour and $T_t$ is the tangential component of $\mathbf T$ along $\partial\Omega$. The contour integral above, expressed in terms of $\mathbf T$, is strictly equivalent to the classical Ampere's law expressed in terms of the magnetic field $\mathbf H$.

In the uncoupled case, each tape turn carries the same current $I_i=I$, where $I$ is equal to source current.
The infinitely thin shell approximation leads to a further simplification of~\eqref{eq:circulation}, because only the component of $\mathbf T$ perpendicular to the tape exists, which nullifies the contour integral everywhere except at the two edges of the conductor. As explained in~\cite{Zhang:SST17,Liang:JAP17}, the desired current $I_i$ flowing in each tape is obtained by imposing the normal components of $\mathbf T$ on the two edges of the tape, i.e.~$T_i^{\rm left}$ and $T_i^{\rm right}$, as Dirichlet boundary conditions for the local $T$ problem in tape $i$. The contour integral in \eqref{eq:circulation} then becomes:
\begin{equation}\label{eq:TA_bc}
I_i=(T_i^{\rm right}-T_i^{\rm left}) \times d,
\end{equation}
where $d$ is the thickness of the superconductor.
Since $T$ is not unique, we can further simplify by setting $T_i^{\rm left}=0$ and $T_i^{\rm right}=T_i$, thus $T_i=I_i/d$, i.e. in the thin sheet approximation, the Dirichlet condition $T_i$ imposes directly the desired current as a sheet current density. The geometric thickness of the tape can thus be ignored in the finite element mesh, making simulations much faster.

In the two coupled cases, the current flowing in the cable turns is known, but the current flowing in each tape turn is not. In order to find the value of $I_i$ in \eqref{eq:TA_bc} for each tape turn, we use an auxiliary electrical circuit, which is available in the AC/DC module of COMSOL Multiphysics~\cite{Wang:JPDAP19}. Therefore, for the two coupled cases, three COMSOL modules are simultaneously used: the magnetic field module for the magnetic field calculation in the whole simulated geometry (including the space outside the superconductors);  the partial differential equations (PDE) module for the calculation of the current vector potential $T$ in the tapes; the electrical circuit module for the calculation of the current in each tape turn.

In order to illustrate the implementation of the circuital method, we consider the minimum working example of a racetrack coil composed of two cable turns, where the cable is made of two stacked tapes, as displayed in Fig.~\ref{fig:Gabriel}(b). We consider only the cross section of the straight part of the coils, i.e. the $xy$ plane in Fig.~\ref{fig:Gabriel}. In addition, due to symmetry, we consider only one of the two straight parts of the coil: in the figure, the one where the tape turns are numbered from 1 to 4.

\begin{figure}[h!]
     \centering
    \includegraphics[width=8 cm]{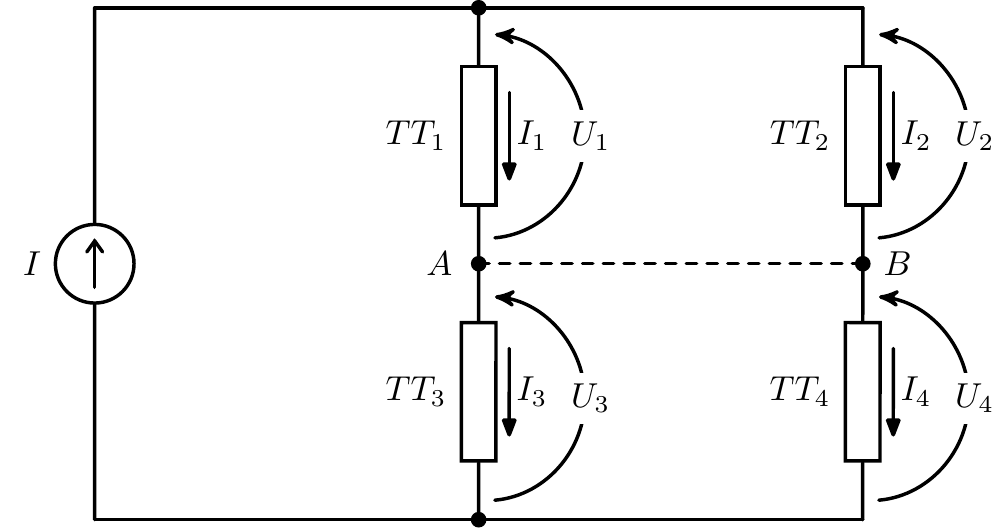}
     \caption{\label{fig:circuito}Electrical circuit for  simulating the cross section of the straight part of an HTS racetrack coil with two cable turns, each composed of two stacked tapes. The full lines represent the connections for coupled-at-ends tape turns. The dashed line represents the additional connection for coupled tape turns.}
\end{figure}

In the circuit, a current source is used to supply the total current $I$ flowing in the cable.
Then, the coupled-at-end and coupled cases differ by the way the tape turns ($TT$) are connected:
\begin{itemize}
\item In the coupled-at-ends case, $TT_1$ and $TT_3$ are physically part of the same tape, so they carry the same (unknown) current, 
that is $I_1=I_3$.
Therefore, in the circuit they are connected in series. The same thing applies to $TT_2$ and $TT_4$, with $I_2=I_4$ (see Fig.~\ref{fig:circuito}).
\item In the coupled case, the tape turns inside each cable are in electrical contact. This means that $TT_1$ and $TT_2$ are in parallel, $U_1=U_2$.
The same applies to $TT_3$ and $TT_4$, i.e. $U_3=U_4$.
In the circuit, this condition is obtained by short-circuiting nodes $A$ and $B$,
as shown by the dashed line in Fig.~\ref{fig:circuito}.
\end{itemize}

In the electrical circuit of COMSOL Multiphysics, each tape turn is defined as an {\it External I vs U} element, with an electric potential at its ends.
For each tape turn, the electric potential is defined as the average of $E+\partial A/\partial t$ (multiplied by a length, here chosen equal to \SI{1}{\meter}), where $E$ and $A$ are the electric field and magnetic vector potential variables.

In the PDE module for $T$, the boundary condition for the vector potential that defines the current flowing in each tape turn (according to~\eqref{eq:TA_bc}) is expressed by means of the current variable calculated in  the electrical circuit.

\section{Results}

In order to compare the three coupling scenarios, we chose the coil configuration presented in~\cite{Grilli:TAS21}. Analyzed is the 2D cross section of the straight part of a racetrack coil, composed of four cable turns, each made of 16 stacked tapes.\footnote{In the text of~\cite{Grilli:TAS21}, it is erroneously indicated that the stacked tapes are 13 instead of 16.} 

The superconductor is modeled as a material with a power-law resistivity
\begin{equation}
\rho(J)=\frac{E_{\rm c}}{J_{\rm c}}\left | \frac{J}{J_{\rm c}} \right |^{n-1},
\end{equation}
where $E_{\rm c}$ is the critical electric field, $J_{\rm c}$ is the critical current density, and $n$ the power-law exponent defining the steepness of the $E$-$J$ curve.
The peak transport current of each cable is $I$=\SI{2248.6}{\ampere} at a frequency of \SI{500}{\hertz}. Different current ratios are obtained by changing the critical current $I_{\rm c}$.
In this work, $J_{\rm c}$ is considered constant in order to simplify the comparison between the different coupling scenarios, but the field dependence can be easily inserted~\cite{Benkel:TAS20}.

\begin{figure}[h!]
     \centering
    \includegraphics[width=8 cm]{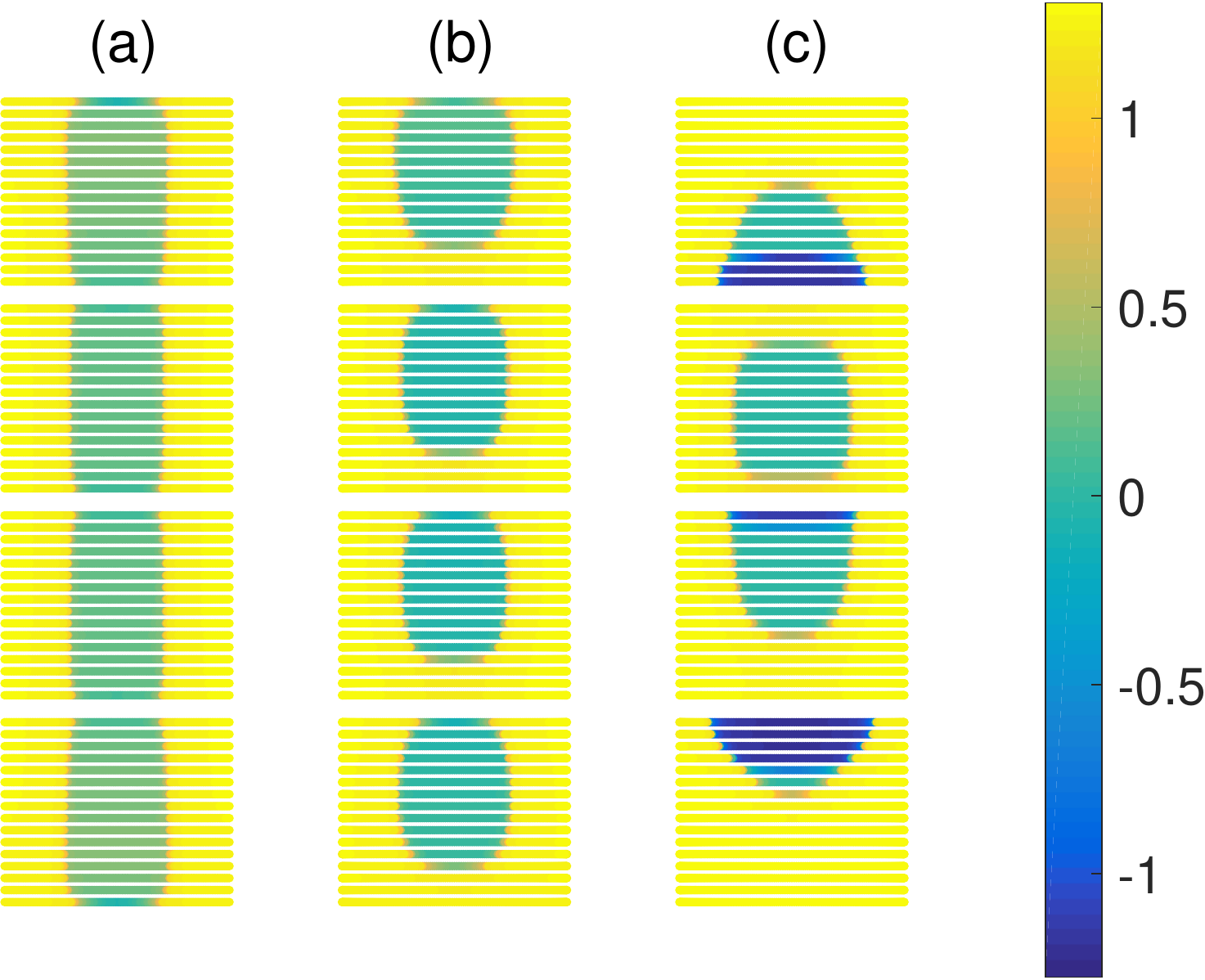}
     \caption{\label{fig:J_distributions}Normalized current density ($J/J_{\rm c}$) distributions for the three coupling scenarios: (a) uncoupled, (b) coupled-at-ends, and (c) coupled tape turns. The transport current is \SI{80}{\percent} of the critical current of the cable turns. Due to symmetry, only one half of the coil is shown.}
\end{figure}

Fig.~\ref{fig:J_distributions} shows the normalized current density ($J/J_{\rm c}$) distributions for the three coupling scenarios: (a) uncoupled, (b) coupled-at-ends, and (c) coupled tape turns. The transport current is \SI{80}{\percent} of the critical current of the cable turns. In the uncoupled case, this means that each tape is also carrying \SI{80}{\percent} of its critical current.

In the uncoupled case (a), the current density distribution in all tapes is similar. None of the tapes is saturated with current, which is beneficial for the AC losses. In the coupled case (c), the situation is very different, with large portions of the cable (i.e. several tape turns) that are completely saturated with current. As shown later, this is a negative situation from the point of view of the power dissipation. Certain regions of the cable turns are also characterized by magnetization currents, as demonstrated by the presence of current densities of opposite sign.
In the coupled-at-ends case (b), the situation is somehow between the previous two, with only some tapes that are fully saturated with current. 

\begin{figure}[h!]
     \centering
    \includegraphics[width=8 cm]{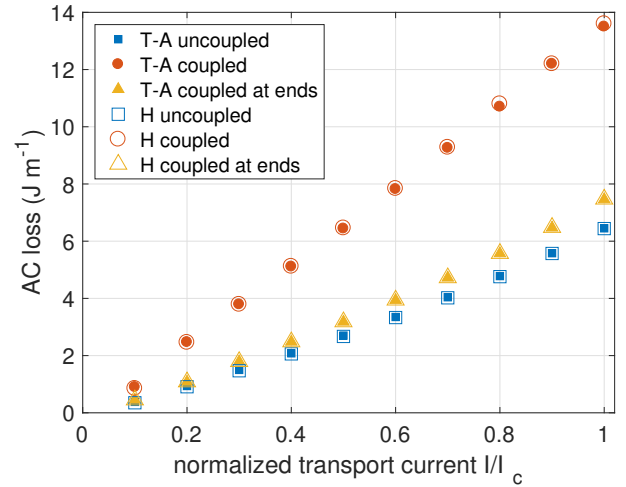}
     \caption{\label{fig:AC_loss}AC losses as a function of the current ratio for the three coupling scenarios: comparison between $T$-$A$ and $H$ formulations.}
\end{figure}

The differences in the current density distribution are mirrored by the level of power dissipation, which is reported in Fig.~\ref{fig:AC_loss} as a function of the current ratio. The uncoupled case is the one with the lowest losses, due to the lack of current saturation of the tapes. The coupled-at-ends case has slightly higher losses due to the saturation of some tapes. The coupled case is the one with the highest losses (exceeding a factor of two with respect to the uncoupled case), due to the very non-uniform current distribution between the tape turns of each cable and the large areas saturated with current.
The figure also shows the results obtained with the widely used $H$ formulation, which has been thoroughly validated against experimental results and other numerical models by several groups~\cite{Shen:SST20,Shen:Access20} and is used here for validation purposes. With the $H$ formulation, the use of current constraints makes the implementation of the coupling scenarios quite straightforward~\cite{Brambilla:SST07}. 
These results confirm the findings presented in~\cite{Pardo:TAS19}, with the coupled-at-ends case having similar losses as the uncoupled case.

\section{Conclusion}
This article describes how to use the $T$-$A$ formulation for modeling HTS coils in 2D, pointing out the  conditions to simulate different coupling scenarios between the tape turns. In particular, the use of an additional electrical circuit module is introduced to simulate cable turns made of tapes that are electrically either along their entire length or just at the ends.
The same numerical models, presented here for stand-alone HTS coils, can be implemented in models for whole electrical machines, and provide a convenient numerical tool for estimating the power dissipation of electrical machines employing HTS materials.


\end{document}